\documentclass[12pt,eqsecnum,preprint]{aastex}

\begin{document}

\title{The Discovery of Soft X-ray Loud Broad Absorption Line Quasars}
\author{Kajal K. Ghosh\altaffilmark{1}} \and\author{Brian
Punsly\altaffilmark{2}} \altaffiltext{1}{Universities Space Research
Association, NASA Marshall Space Flight Center, VP62, Huntsville,
AL, USA} \altaffiltext{2}{4014 Emerald Street No.116, Torrance CA,
USA 90503 and ICRANet, Piazza della Repubblica 10 Pescara 65100,
Italy, brian.m.punsly@L-3com.com or brian.punsly@gte.net}

\begin{abstract}It is been known for more than a decade that
BALQSOs (broad absorption line quasars) are highly attenuated in the
X-ray regime compared to other quasars, especially in the soft band
( $< $ 1 keV). Using X-ray selection techniques we have found ``soft
X-ray loud" BALQSOs that, by definition, have soft X-ray (0.3 keV)
to UV ($3000 \AA$) flux density ratios that are higher than typical
nonBAL radio quiet quasars. Our sample of 3 sources includes one
LoBALQSO (low ionization BALQSO) which are generally considered to
be the most highly attenuated in the X-rays. The three QSOs are the
only known BALQSOs that have X-ray observations that are consistent
with no intrinsic soft X-ray absorption. The existence of a large
X-ray luminosity and the hard ionizing continuum that it presents to
potential UV absorption gas is in conflict with the ionization
states that are conducive to line driving forces within BAL winds
(especially for the LoBALs).
\end{abstract}


\section{Introduction} One of the biggest challenges to our
understanding of broad absorption line quasars (BALQSOs, hereafter)
is how lithium like species (those species producing the resonant UV
absorption lines) can form within an out-flowing wind in the
presence of the hard ionizing continuum of a quasar (EUV and X-ray).
Thus, it was no surprise that BALQSOs turned out to be X-ray weak
due to thick absorption columns \citep{gre96}. The necessity of this
X-ray absorption screen between the central quasar X-ray source and
the BAL wind was anticipated by \citet{mur95}. Every deep X-ray
observation of a BALQSO has shown significant absorption with the
least absorbed sources having neutral hydrogen absorption columns
$N_{H} \gtrsim 10^{22} \mathrm{cm}^{-2}$ and most BALQSOs have
absorption columns of $ 10^{23} \mathrm{cm}^{-2}\leq N_{H} \leq
10^{25} \mathrm{cm}^{-2}$ \citep{pun06}. As a qualifier, it is
possible that the modest radio jet seen in some BALQSOs can also
contribute to the X-ray flux and some of these sources might not
appear X-ray suppressed due to a secondary source of X-rays from the
relativistic jet $\sim$ 10 - 100 pc from the central quasar
\citep{bro05}. Thus, we make a distinction between radio quiet
(RQBALQSOs), $\log R<1$, and radio loud BALQSOs (RLBALQSOs), $\log
R>1$, where R is the k-corrected 5 GHz to $2500\AA$ flux density
ratio (the RQBALQSOs considered here have $\log R<0$). The few
RQBALQSOs with modest absorption, UM425 and CSO755, $N_{H} \gtrsim
10^{22} \mathrm{cm}^{-2}$ have fairly typical hard X-ray to UV flux
density ratios for radio quiet quasars, but are highly attenuated in
the soft X-rays \citep{ald03,she05,mur95}. To this date, all known
RQBALQSOs are highly absorbed in the soft X-rays. The lone exception
is the X-ray spectrum of IRAS 07598+059 that is consistent with no
absorption. However, the X-ray flux relative to the UV is suppressed
by a factor of more than 20 compared to a typical radio quiet
quasar, thus it is believed that the central quasar X-ray source is
completely obscured and we are seeing a secondary weak source,
likely the central source seen in reflection off an electron
scattering mirror \citep{ima04}.
\par We have searched the ROSAT database to look for evidence of soft
X-ray loud BALQSOs. Previously, the only BALQSO ROSAT detection in
the literature is 1245-067 and it is highly absorbed, $N_{H} \approx
10^{23} \mathrm{cm}^{-2}$ \citep{gre96}. We have found approximately
40 likely detections of BALQSOs by ROSAT and are preparing a catalog
of these. However, most of these either are not necessarily strong
compared to the UV flux or have poor photon statistics, so it is
difficult to determine the X-ray luminosity with much certainty. In
sorting through the catalog, we have segregated three AGN that are
the most convincing examples of RQBALQSOs that are soft X-ray loud
(defined by soft X-ray (0.3 keV) to UV ($3000 \AA$) flux density
ratios that are higher than typical nonBAL radio quiet quasars), one
of which has low ionization UV absorption lines (LoBALQSO). This is
even more surprising since the the LoBALQSOs require the maximal
amount of X-ray screening, otherwise the gas will be over-ionized
for the formation of Li-like low ionization species \citep{mur95}.
These BALQSOs could either be intrinsically exceptional objects or
objects that are fortuitously configured to reveal some unforseen
physical features of the QSO central engine.
\section{The X-ray and UV Data}
A search for optical counterparts in the Sloan Digital Sky Survey
Data Release 4 of ROSAT PSPC sources from the White-Giommi-Angelini
Catalog, \citet{whi94}, was performed by \citet{suc06}. They found
1744 tentative identifications of ROSAT sources with optical spectra
typical of AGN. Our in depth inspection of  the SDSS spectra of
these 1744 AGNs has identified approximately 44 AGNs as BALQSOs.
This section describes the data for the three candidate BALQSOs that
are soft X-ray loud, SDSS J023219.52+002106.8, SDSS
J101949.75+450256.0 and SDSS J132229.66+141007.3 (corresponding to
the WGA counterparts 1WGA J0232.3+0021, 1WGA J1019.7+4502 and 1WGA
J1322.4+1410, respectively). Hereafter, the SDSS designations will
be abbreviated as 0232, 1019 and 1322.
\begin{figure*}
\includegraphics[width=50 mm, angle= -90]{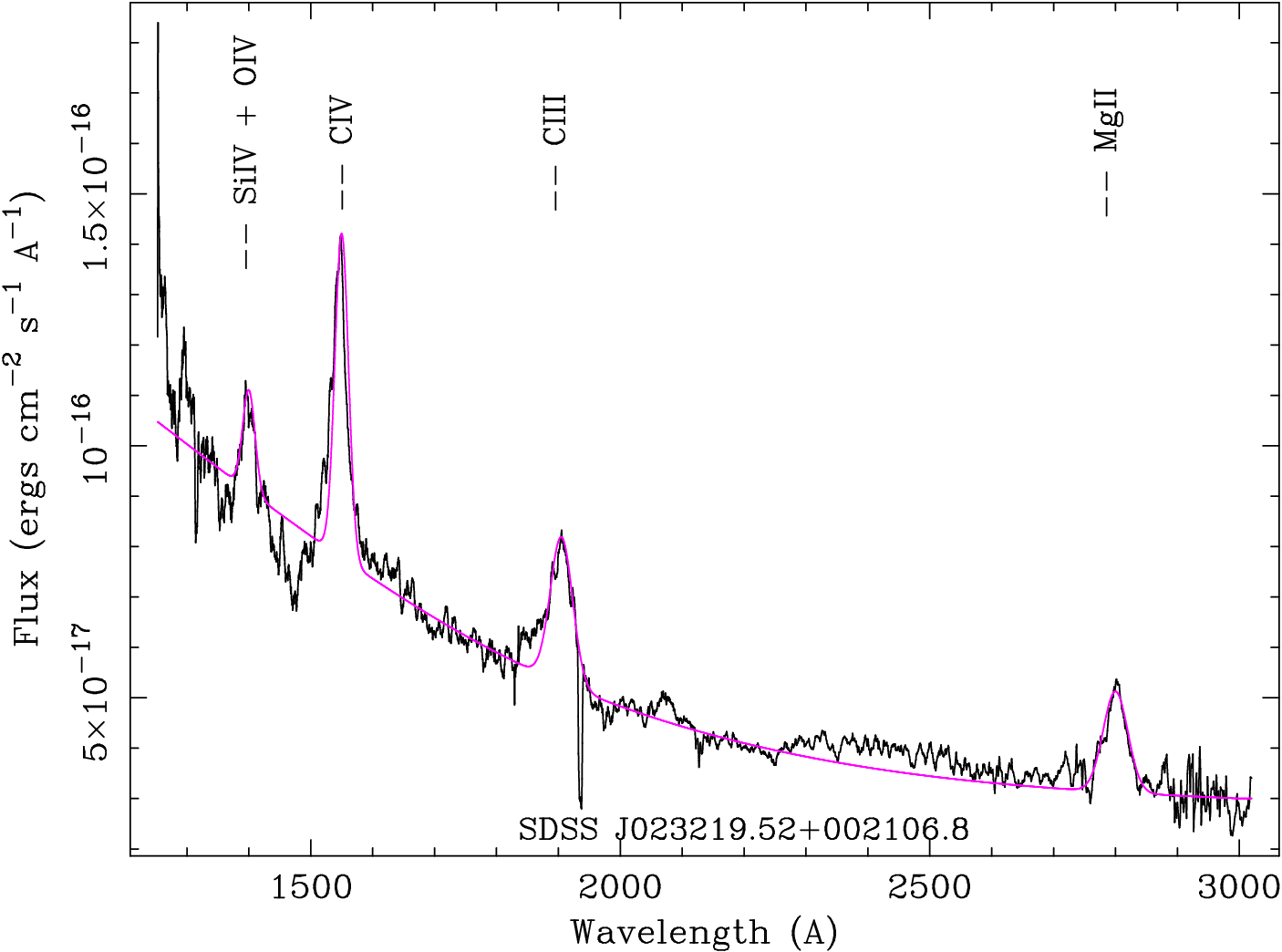}
\includegraphics[width=50 mm, angle= -90]{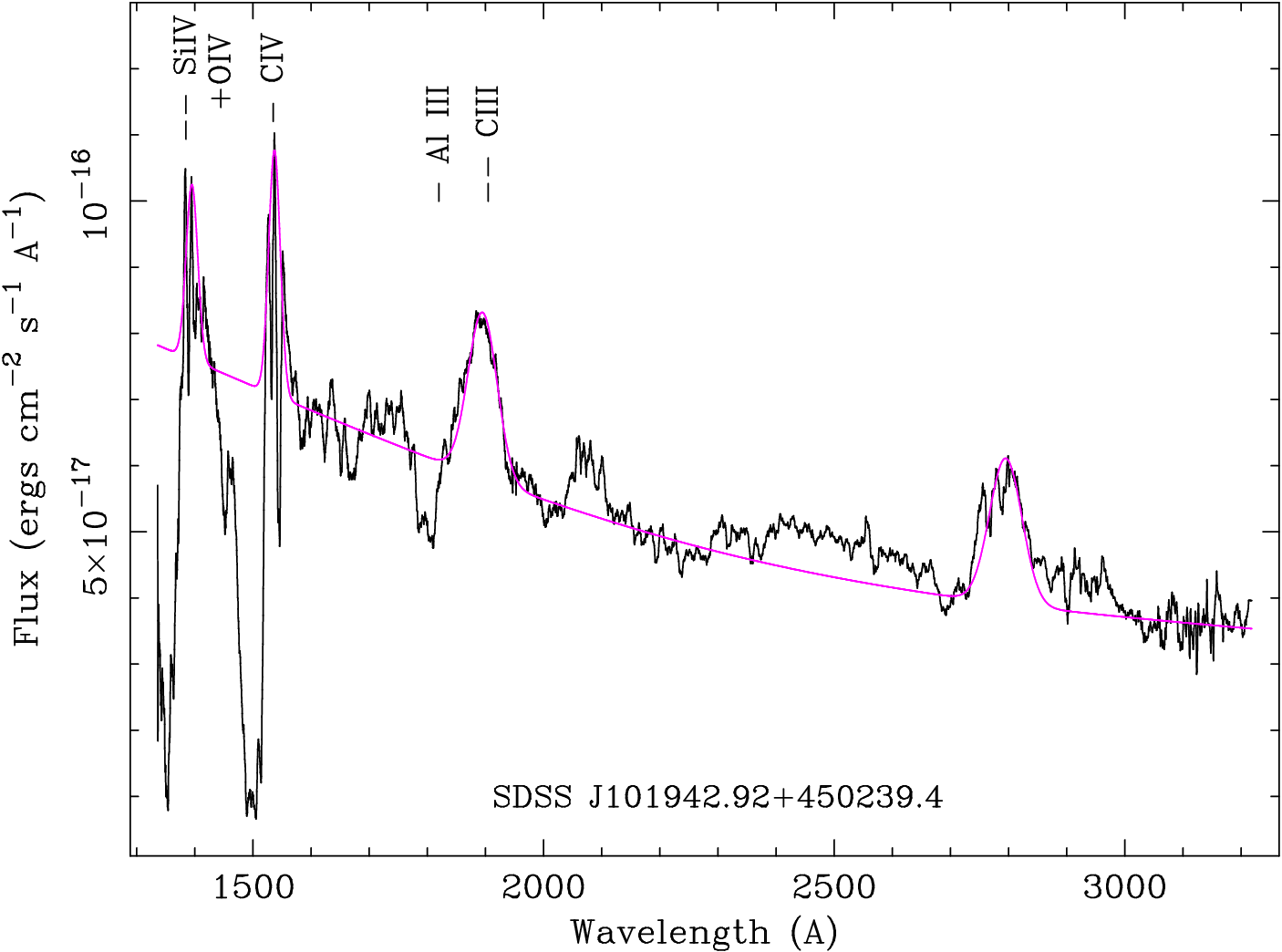}
\includegraphics[width=50 mm, angle= -90]{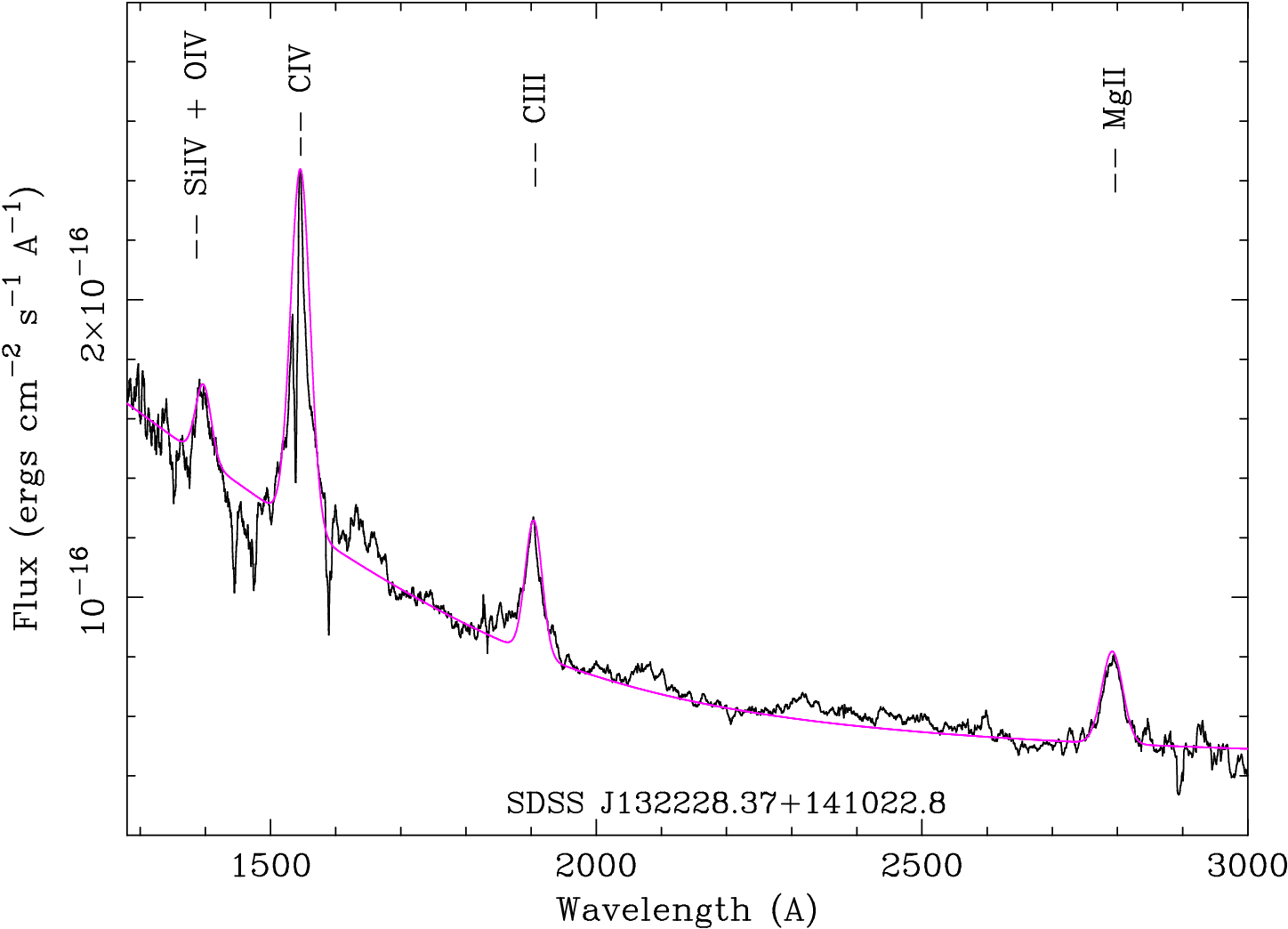}
\caption{The UV spectra from SDSS (from left to right, 0232, 1019, 1322), the best fit continuum is in magenta.}
\end{figure*}
\par Event files for the ROSAT PSPC observations of 0232, 1019 and 1322 were  retrieved from the
ROSAT archive. We use the locally-developed software tool, LEXTRCT,
for the extraction of source and background light curves and spectra
\citep{ten06}. Good time interval filtering was applied to all the
event files. We extracted source counts from circles of 60\arcsec\
radius around each source and corresponding backgrounds from nearby
source-free regions. Background regions were around 3 to 5 times
larger than the source regions. X-ray light curves were binned into
1000~s bins. X-ray spectra of all three sources were binned so that
there were at least 30 counts per fitting bin. Spectral
redistribution matrices and ancillary response files were retrieved
from the ROSAT archive. XSPEC version 11.3 was used to fit the 0.2
-- 2.4 keV energy spectra with an absorbed ({\tt phabs}) powerlaw
({\tt powerlaw}) model. Confidence contours in
the $\Gamma$ (photon index)--$N_{H}$ plane were generated for the models. The confidence contours
indicate that $N_{H} < 2 \times 10^{21}\mathrm{cm}^{-2}$ with
$>99\%$ confidence. The photon statistics are not good enough to
segregate out the small intrinsic absorption column density from the
Galactic value of $N_{H}$. Thus, $N_{H}$ in of our favored model fit (column
6 of Table 1) is fixed at the Galactic value. We also fitted these spectra with a
redshifted, absorbed, powerlaw model. Best fitting model parameters
are similar to the previous values. Table 1 describes the details of
our fits to the data. The first two columns are the source name and
redshift. The next column is X-ray luminosity at the rest frame of
the QSOs in the 2-8 keV energy band. $\Gamma$ is tabulated in
columns 4 and 5 based on $\chi^{2}$ and Cash statistics,
respectively. The ``goodness of fit" appears in column 7. In the
following columns, we define two measures of the X-ray flux density
strength relative to the UV: a UV to soft X-ray spectral index
defined in terms of the flux density, $F_{\nu}$ as $\alpha_{os}
\equiv 0.537 \log [F_{\nu}(3000 \AA)/F_{\nu}(0.3 \mathrm{keV})]$,
\citet{lao97}, and the standard UV to hard X-ray spectral index,
$\alpha_{ox} \equiv 0.384 \log [F_{\nu}(2500 \AA)/F_{\nu}(2.0
\mathrm{keV})]$ \citep{gre96}. If there is small nonzero intrinsic
absorption as suggested by the confidence contours in
the $\Gamma$ --$N_{H}$ plane, the intrinsic X-ray luminosity
is even larger than indicated in column 3 (i.e., $\alpha_{os}$ and
$\alpha_{ox}$ are even flatter than in columns (8) and (9)). Thus,
our claims that these sources are soft X-ray loud are conservative.
Deeper X-ray observations are required to improve the accuracy of
our spectral fits.
\begin{table}
\caption{X-ray Spectral Models and Balnicity Indices of Soft X-ray
Loud BAL Quasars} {\tiny
\begin{tabular}{cccccccccccc} \tableline \rule{0mm}{3mm}

 QSO &   z & $L_{x}$ & $\Gamma$ & $\Gamma$ &  $N_{H}$  & $\chi^{2}$/dof & $\alpha_{os}$& $\alpha_{ox}$&BI CIV&BI AlIII & Type \\
       &    &  \tablenotemark{a} & $\chi^{2}$ & Cash   &\tablenotemark{b} &   &  &&(km/s)& (km/s)& \\
\tableline \rule{0mm}{3mm}
 0232 & 2.04 & $0.99     $ & $2.61 $ & $2.55 $     & 2.73    & 2.2/6& $1.22^{+0.18}_{-0.17}$ & $1.35^{+0.03}_{-0.02}$ &1626& ... &  HiBAL \\
   &      & $ \pm 0.17 $ & $ \pm 0.35$& $ \pm 0.16$ &   & &&     &&&   \\
  &&&&&&&&&&& \\
 1019 & 1.87 & $ 0.27     $ & $2.89$ & $2.62$  &  1.07                   &0.9/4& $1.33^{+0.30}_{-0.26}$& $1.53^{+0.04}_{-0.02}$ &24135&4032& LoBAL \\
   &      & $ \pm 0.08 $ &$ \pm 0.58$& $ \pm 0.18$  &       & &      &     &&&       \\
  &&&&&&&&&&& \\
 1322 & 2.05 & $ 1.8    $ & $2.62$ & $3.03$  &  1.76                &3.9/6&  $1.16^{+0.25}_{-0.21}$ & $1.33^{+0.03}_{-0.02}$ &1819& ... & HiBAL \\
  &      & $ \pm 0.32 $ &$ \pm 0.49 $ & $ \pm 0.20 $  &     &   &    &      &&&        \\
\end{tabular}}
\tablenotetext{a}{X-ray luminosity derived from an absorbed
powerlaw model in the QSO rest frame from 2.0 -8.0 keV, in units of
$10^{45}$ ergs/s} \tablenotetext{b}{Neutral hydrogen absorption
column used in the model in units of $10^{20}\mathrm{cm}^{-2}$}
\end{table}
\begin{figure*}
\includegraphics[width=50 mm, angle= -90]{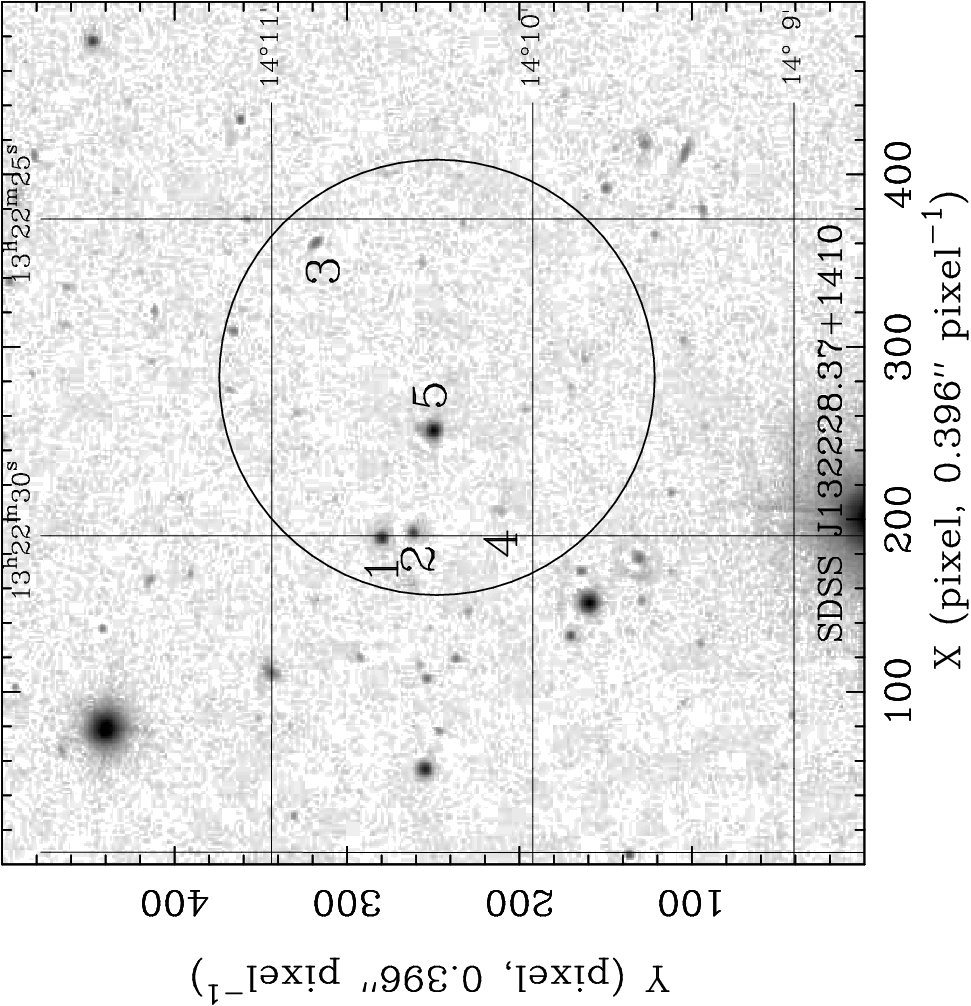}
\includegraphics[width= 50 mm, angle= -90]{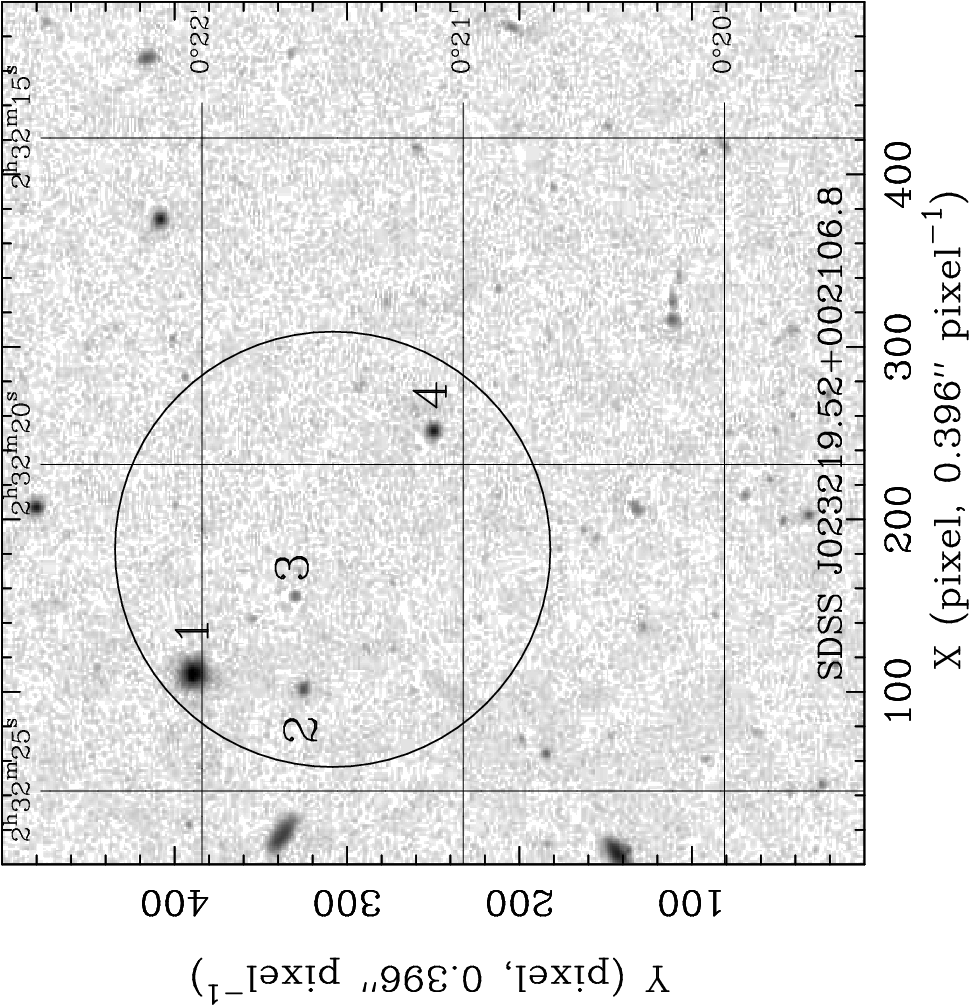}
\includegraphics[width=50 mm, angle= -90]{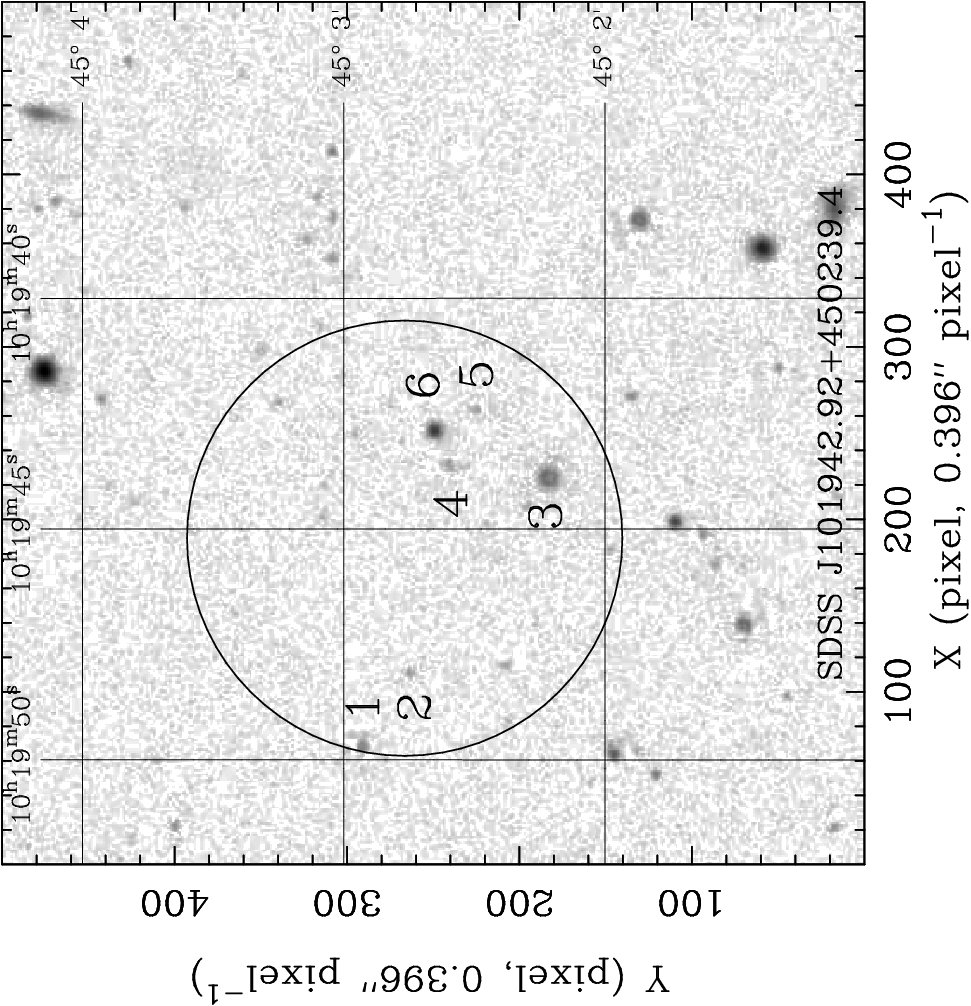}
\caption{Finding charts for the 3 ROSAT detections, from left to
right, 0232, 1019 and 1322. The SDSS identification from
\cite{suc06} is at the center of each frame. The circles represent
the 90\% ROSAT confidence source error radius.}
\end{figure*}
\par The SDSS spectra of the 3 soft X-ray loud RQBALQSOs in the
regions of the BALs are presented in Figure 1. The spectra were retrieved from the SDSS database and were analyzed
using the $IRAF$\footnote{IRAF is the Image Reduction and Analysis
Facility, written and supported by the IRAF programming group at the
National Optical Astronomy Observatories (NOAO) in Tucson, Arizona.}
software. First, the spectrum was de-reddened using the Galactic
extinction curve, \citet{sch98}, then the wavelength scale was
transformed from the observed to the source frame. The spectra were
fitted in XSPEC with a powerlaw plus multiple gaussian model,
including fits for both emission lines and absorption lines
\citep{arn96}. All the model parameters were kept free. The best fit
to the SDSS data was determined using $\chi ^{2}$ minimization. If
an emission bump around 2500 $\AA$ is present then the continuum fit
in the CIV (1550 $\AA$) region was extrapolated to longer
wavelength. The BALnicity indices quoted in Table 1 are a
consequence of the method of spectral fitting described above and
other methods might produce different results. However, the exact
BALnicity index is not critical to this discussion, the spectra in
Figure 1 clearly show BALs and that is the essential point of
relevance here. In columns 10 and 11 of Table 1, we compute the
BALnicity indices (BIs) for the high ionization CIV trough and the
low ionization Al III trough, respectively, per the methods of
\citet{wey91}; BI $>0$ means the source is a BALQSO.
\section{Source Identification}
Since these objects are so exceptional and the ROSAT error circles
are so large, it is important to verify the identifications with the
SDSS AGN by \citet{suc06}. In Figure 2, we present SDSS ``finding
charts" for the 3 ROSAT detections. All of the SDSS sources within
the 90\% confidence source error radius are labeled. Verification of
the identifications in \citet{suc06} is predicated on the individual
probabilities that each source within the confidence source error
radius is not the ROSAT source: the last column in Table 2. In order
to arrive at these probabilities, the table first lists the source
classification (column 2) for each of the sources in the three ROSAT
confidence error radii in Figure 2. The source classifications in
Table 2 are based on SDSS photometry using the results of
\citet{new99,str01} and spectra when available. The classifications
are amended by the SDSS values of $m_{B}$ and the VLA/FIRST radio
flux densities in the next two columns. This allows a comparison of
each putative ROSAT source with the appropriate ROSAT ``standard
candle" that is identified in column 5. Using the source
classification and the hypothesis that each source is the ROSAT
detection, each source can be compared individually to the
appropriate ``standard candle" X-ray properties in columns 6 - 8 in
order to assess whether it is a viable identification. The standard
candle values are listed on top and the putative source values are
listed below in bold and in parenthesis to clearly distinguish the
two. For example, consider the F-star that is source 1 in the field
of 1WGAJ0232.3+0021 in Figure 2 (row 1 of Table 2). Given the value
of $M_{V}$ associated with its stellar classification and $m_{V}$
from SDSS photometry, one can determine the distance to the star. If
one assumes that this is the ROSAT source then one can compute the
intrinsic X-ray luminosity in the ROSAT band, $\log (L_{X})=
34.20\pm 0.20$ which is much larger than any of the ROSAT F stars of
similar $M_{V}$ in \citet{suc03}, $\log (L_{X})= 29.38\pm 0.50$. If
the \citet{suc03} data fits a log normal distribution then we can
rule out the F-star as the ROSAT source with $>0.9999$ probability.
If however, the tail of the probability distribution for $\log
(L_{X})$ of ROSAT F-stars beyond 3 $\sigma$, does not obey Gaussian
statistics (i.e., there are extra outliers in the tail), the F-star
can not be ruled out categorically. The M-stars are treated in a
similar manner using the ``standard candle" $\log (L_{X})$ values
for M-stars of matched $M_{V}$ from \citet{mar00}.

\begin{table}
\caption{Probability of False Identification of the Sources in the
1WGA Fields } {\tiny
\begin{tabular}{ccccccccc} \tableline \rule{0mm}{3mm}

Source &   Type & $m_{B}$ & 1.4 GHz & Standard Candle & & Standard Candle & & Prob.  \\
 &    &  & (mJy) &  &  & (\textbf{Putative Source})  &  & False ID \\
 & & &\tablenotemark{a} & & Flux \tablenotemark{b} & Log $L_{x}$\tablenotemark{c} & $\alpha_{os}$ \tablenotemark{d}& \\
\tableline \rule{0mm}{3mm}
&&&&\textbf{1WGAJ0232.3+0021}&&&&\\
\tableline \rule{0mm}{3mm}
 source 1 & FV7-9  & $16\pm 0.03$ & $<0.99$ & ROSAT F stars,  & ...  & $29.38\pm 0.5$ & ... & $>0.9999$  \\
   &   star   &  & & $3.0 < M_{V} < 4.0$ &  &\textbf($\mathbf{34.20\pm 0.20}$) &&        \\
source 2 & spiral & $20.8\pm 0.1$ & $<0.99$ & radio quiet   & $4.4 \pm 0.7$  & ... & ... & $>0.9999$  \\
   &   galaxy  &  & & spirals & \textbf{(6592)} & &&        \\
source 3 & quasar \tablenotemark{e} & $22.0 \pm 0.15$ & $<0.99$ & radio quiet  & ...  & ... & $1.52 \pm 0.26$ & $>0.9999$  \\
   &   $z=0.768$   &  & & unobscured quasar &  & &\textbf{($\mathbf{0.56^{+0.14}_{-0.13}}$)}&        \\
source 4 & BALQSO  & $19.4\pm 0.05$ & $<0.99$ & radio quiet  & ...  & ... & $1.52 \pm 0.26$  & $0.8289$  \\
   &      &  & & quasar &  &... &\textbf{($\mathbf{1.22^{+0.18}_{-0.17}}$)}&        \\
\tableline \rule{0mm}{3mm}
&&&&\textbf{1WGAJ1019.7+4502}&&&&\\
\tableline \rule{0mm}{3mm}
source 1 & M3.5 - M4.5  & $20.7\pm 0.1$ & $<0.98$ & ROSAT M3.5   & ...  & $27.72\pm 1.05$ & ... & $0.9982$  \\
   &   star   &  & & - M4.5 stars &  &\textbf($\mathbf{31.15\pm 0.54}$) &&        \\
source 2 & spiral & $22.9\pm 0.3$ & $<0.98$ & radio quiet   & $0.63 \pm 0.7$  & ... & ... & $>0.9999$  \\
   &   galaxy  &  & & spirals & \textbf{(3217)} & &&        \\
source 3 & M4.5 - M5.5  & $17.6\pm 0.03$ & $<0.98$ & ROSAT M4.5  & ...  & $26.96\pm 0.61$ & ... & $>0.9999$  \\
   &   star   &  & & - M5.5 stars &  &\textbf($\mathbf{29.84\pm 0.10}$) &&        \\
source 4 & spiral & $21.3\pm 0.1$ & $<0.98$ & radio quiet   & $2.8 \pm 0.7$  & ... & ... & $>0.9999$  \\
   &   galaxy  &  & & spirals & \textbf{(3217)} & &&        \\
source 5 & elliptical & $22.7\pm 0.25$ & $<0.98$ & radio quiet   & $3.1 \pm 0.7$  & ... & ... & $>0.9999$  \\
   &   galaxy  &  & & ellipticals & \textbf{(3217)} & &&        \\
source 6 & BALQSO  & $19.3\pm 0.05$ & $<0.98$ & radio quiet  & ...  & ... & $1.52 \pm 0.26$  & $0.7675$  \\
   &      &  & & quasar &  &... &\textbf{($\mathbf{1.33^{+0.30}_{-0.26}}$)}&        \\
\tableline \rule{0mm}{3mm}
&&&&\textbf{1WGAJ1322.4+1410}&&&&\\
\tableline \rule{0mm}{3mm}
source 1 & FV8-9  & $19.4\pm 0.05$ & $<0.91$ & ROSAT F stars,  & ...  & $29.38\pm 0.5$ & ... & $>0.9999$  \\
   &   star   &  & & $3.0 < M_{V} < 4.0$ &  &\textbf($\mathbf{34.11\pm 0.20}$) &&        \\
source 2 & FV6-7  & $20.6\pm 0.1$ & $<0.91$ & ROSAT F stars,  & ...  & $29.38\pm 0.5$ & ... & $>0.9999$  \\
   &   star   &  & & $3.0 < M_{V} < 4.0$ &  &\textbf($\mathbf{34.76\pm 0.53}$) &&        \\
source 3 & spiral & $21.3\pm 0.1$ & $<0.91$ & radio quiet   & $3.8 \pm 0.7$  & ... & ... & $>0.9999$  \\
   &   galaxy  &  & & spirals & \textbf{(12046)} & &&        \\
source 4 & elliptical & $22.6 \pm 0.2$ & $<0.91$ & radio quiet   & $3.1 \pm 0.7$  & ... & ... & $>0.9999$  \\
   &   galaxy  &  & & ellipticals & \textbf{(12046)} & &&        \\
source 5 & BALQSO  & $18.4\pm 0.03$ & $<0.91$ & radio quiet  & ...  & ... & $1.52 \pm 0.26$  & $0.8413$  \\
   &      &  & & quasar &  &... &\textbf{($\mathbf{1.16^{+0.25}_{-0.21}}$)}&        \\
\end{tabular}}
\tablenotetext{a}{\tiny{Upper limit on flux density from VLA/FIRST}}
 \tablenotetext{b}{\tiny{X-ray flux in the
observers frame 0.2 - 2.4 keV ($10^{-17}$ ergs/s
$\mathrm{cm}^{-2}$)}} \tablenotetext{c}{\tiny{Putative source
luminosity is computed from $\chi^{2}$ fit to ROSAT data in figure
1}} \tablenotetext{d}{\tiny{$\alpha_{os}$ is defined in
\citet{lao97} and the characteristic quasar range of values is from
that paper}}\tablenotetext{e}{\tiny{This is not an obscured (red
quasar) based on SDSS colors and 2MASS upper limit on flux density
that yields, $R-K< -0.8$. The photometric redshift is based on SDSS
colors}}
\end{table}
Similarly, there is a standard candle for each field galaxy in
Figure 3 that is determined by the galaxy type, the 1.4 GHz flux
density and $m_{B}$. The X-ray fluxes of radio quiet galaxy standard
candles are given by \citet{ref97}. These standard candle fluxes are
compared to the observed ROSAT fluxes in the 1WGA fields in column
6. The ROSAT fluxes in the 1WGA fields are three orders of magnitude
too large to be consistent radio quiet galaxy counterparts. The
quasar soft X-ray properties have been studied in \citet{lao97} by
means of $\alpha_{os}$. The quasar values of $\alpha_{os}$ are
compared to the \citet{lao97} characteristic quasar values in column
(8). The only objects in Table 2 without a high statistical
significance for not being the ROSAT detection are the BALQSOs.
\par There is still the possibility that the 1WGA sources are interlopers,
i.e. it was anomalously bright during the PSPC observations and then
faded subsequently. The WGACAT catalog contains more than 68000
sources and a total of 216 variable sources detected that could be
considered interlopers \citep{whi94}. Thus, the chance of finding an
interloper in one of these 3 fields is $<0.0095$.
\section{Discussion}
In this Letter, we describe 3 BALQSOs that have soft X-ray to UV
ratios that are larger than typical quasars (see the $\alpha_{os}$
values in Table 2). Hence, the designation as soft X-ray loud.
However, the $\alpha_{ox}$ values are not exceptional for quasars
due to the steep X-ray spectral indices \citep{str05}. These are
also the first known BALQSOs with X-ray absorption consistent with
pure Galactic absorption. The tentative discovery of soft X-ray loud
BALQSOs is completely unexpected based on theoretical treatments of
BALQSOs. The conundrum posed by this new class of AGN is how can a
BAL region coexist with a powerful source of X-rays, since even a
modest flux of X-rays will over-ionize the gas making it impossible
to form Li-like atoms \cite{mur95}. On a speculative note, a
relativistic X-ray jet beamed towards earth and away from the BAL
gas is a possibility \citep{pun99,pun00}. It is now known that there
are BALQSOs with relativistic jets beamed toward earth and a jet
similar to the one in Figure 1 of \cite{gho07} that is radio weak,
but X-ray strong would conform to the properties of the BALQSOs in
Tables 1 and 2. There are extragalactic jets that have X-ray and
radio properties in-line with theses sources. Extreme high frequency
peaked BL Lac objects such as PG1553+113, at $z \approx 2$, would
have an apparent X-ray luminosity of $\gtrsim 10^{45}$ ergs/s, a
steep X-ray spectrum and would have a radio flux at 1.4 GHz of about
0.1 - 0.5 mJy in agreement with the properties of the BALQSOs
presented here \citep{ost06}.

\end{document}